\documentstyle[aps]{revtex}
\newcommand{\be}{\begin{equation}}
\newcommand{\ee}{\end{equation}}
\begin{document}
\draft
\title{GROUP ANALYSIS AND RENORMGROUP SYMMETRIES}
\author{VLADIMIR~F.~KOVALEV}
\address{Institute for Mathematical Modelling of Russian Academy
          of Sciences, \\
          Miusskaya 4-A, Moscow, 125047, RUSSIA
          E-mail:kovalev{\#}9@imamod.msk.su }
\author{\fbox{VENIAMIN~V.~PUSTOVALOV} }
\address{ P.N.Lebedev Physical Institute,
          Leninskii pr.,53, Moscow, 117924, RUSSIA}
\author{DMITRII~V.~SHIRKOV}
\address{ Bogoliubov Lab., Joint Institute for Nuclear Research, \\
          Dubna, 141980, RUSSIA
          E-mail: shirkovd@thsun1.jinr.dubna.su  }

\maketitle

\vspace{0.2cm}
\begin{flushright}
hep-th/9706056
\end{flushright}

\medskip

\begin{abstract}
An original regular approach to constructing special type symmetries for
boundary value problems, namely renormgroup symmetries, is presented.
Different methods of calculating these symmetries, based on modern group
analysis are described. Application of the approach to boundary value
problems is demonstrated with the help of a simple mathematical model.
\end{abstract}


\makeatletter
\def\@oddhead{\hfill \small{Kovalev, Pustovalov, and Shirkov:
Group analysis and RG-symmetries}  \hfill }
\let\@evenhead\@oddhead

\section{Introduction}
The paper is devoted to the problem of constructing a special class
of symmetries for boundary value problems (BVPs) in mathematical
physics, namely {\bf renormalization group symmetries} (hereafter
referred to as RG-symmetries).
\par
Symmetries of this type appeared about forty years ago in the
context of the renormalization group (RG) concept.  This concept
originally arose \cite{StPt,GL,BSh_DAN_55,BSh_NC_56} in the "depth"
of quantum field theory (QFT) and was connected with a complicated procedure
of renormalization, that is 'removing of ultra-violet infinities'. In QFT
renormalization group was based upon finite Dyson transformations and
appeared as a continuous group in a usual mathematical sense. It was
successfully used for improving approximate perturbation solution to
restore a correct structure of a solution singularity.
\par
In the seventies, it was found that the RG concept was fruitful in some other
fields of microscopic physics: phase transitions in large statistical
systems, polymers, turbulence, and so on.  However, in some cases, following
Wilson's approach \cite{Wilson} to spin lattice, the original {\it exact
symmetry} underlying the renormalization group notion in QFT was changed to an
approximate one with the corresponding transformations forming a semi-group
(not a group as in the QFT case). Here, in this paper, by RG-symmetry we mean
the original exact property of a solution -- as it was formulated in
Refs.\cite{BSh_DAN_55,BSh_NC_56}
(see also \cite{Book}) by N.~Bogoliubov and one of the
present authors. Thus, {\it by RG-symmetry we mean a symmetry that
characterizes a solution of a BVP and corresponds to {\bf transformations}
involving both "dynamical" (i.e., equation) {\bf variables} and {\bf
parameters} entering into a solution via equations and boundary conditions}.
\par
For a simple illustration we consider some BVP that produces a family of
solutions. The simplest variant of RG transformation is given by a
simultaneous one-parameter point transformation
\begin{equation} T_a : \left\{x \to x' = x/a~,~ g \to g'= G(a, g)~\right\}~,
\quad G(1, g)=g
\label{eq1.1} \ee
of a dimensionless "coordinate" $x$ and a one-argument "characteristic"
(e.g., initial value) $g$ of each solution, the quantity of a
direct physical interest.
The transformation function $G(x, g)$, which depends upon two
arguments \cite{ftn1}, should satisfy the functional equation
\be G(x, g) = G(x/a, G(a,g)),
\label{eq1.2} \ee
that guarantees the group property $T_a\cdot T_b = T_{ab}$ fulfillment.
\par
The functional equation (\ref{eq1.2}) and transformation
(\ref{eq1.1}) arise, for example, in the massless QFT with one
coupling. In that case  $x = Q^{2}/\mu^{2}$ is the ratio of a
4-momentum $Q$ squared to a "reference momentum" $\mu $ squared,
$g$ is the coupling constant and $G$ is the so-called {\it
effective coupling}.
\par
Later on the (\ref{eq1.1})-(\ref{eq1.2})--type symmetry underlying the
renormgroup invariance was also found in a number of problems of macroscopic
physics like, e.g., mechanics, transfer theory, hydrodynamics and a close
relation of RG-symmetry to the notion of self-similarity was
established \cite{Sh_SPD_82,Sh_TMF_84}.
\par
The infinitesimal form  of transformation (\ref{eq1.1})
 can be written down as a differential equation
\be R \, G = 0 \,, \quad \mbox{with}
~~ R=x\partial_x - \beta(g)\partial_g \,, \quad
\beta (g) = {\frac{\partial G (a; g)}
            {\partial a}}{\bigg\vert_{a=1}}
\label{eq1.3} \ee
where $R$ is the infinitesimal operator of RG-symmetry
(or, simply, {\it RG-operator}) with the coordinate $\beta(g)$ defined by
the derivative of the function $G$.
\par
Therefore, instead of relations (\ref{eq1.1}) and (\ref{eq1.2}), RG
transformation can also be introduced by means of an RG-operator.
And vice versa, being given an RG-operator one can reconstruct the
functional equation for the solution $G$ with the help of the characteristic
equation for (\ref{eq1.3}).  Moreover, for a given $\beta$-function or, in
other words, for the given RG-symmetry, one can get an explicit expression
for the invariant of the group transformation $G(x,g)$ by solving
the corresponding Lie equations \cite{Ovs82}
\be
- \frac{d x'}{x'} = \frac{dg'}{\beta(g')} = \frac{da}{a},
\label{eq1.5} \ee
with the initial conditions $ \quad x'\vert_{a=1}=x, \ g'\vert_{a=1}=g$.
\par
Along with (\ref{eq1.3}) a different form of the {\it invariance
condition}\/ for the function $G(x,g)$ is often employed
\be
\frac{d}{da}\,G(x/a , G(a,g) ) {\bigg\vert_{a=1}}  = 0\,.
\label{eq1.3'} \ee
Equation (\ref{eq1.3}), reflecting the invariance of $G$ under the RG
transformation can be treated as a vanishing condition for the coordinate
$\mbox{\ae}$ of the RG operator (\ref{eq1.3}) in the canonical form \cite{Ibr}
\be
\bar R=\mbox{\ae}\partial_{G}\,, \quad
\mbox{\ae} \equiv x {G}_x - \beta(g) {G}_g = 0 \,,
\label{eq1.4} \ee
identically valid on a particular BVP solution $G = G (x,g)$.
\par
At the same time, the relation
\be  R S(x, g) \equiv \left( x \partial_x - \beta (g)
\partial_g \right) S(x, g)= \gamma (g) S(x, g)
\label{eq1.4'} \ee
corresponds to the function $S(x, g)$ that is a {\it
covariant} \cite{ftn2} of the RG transformation. In QFT case,
this relates, e.g., to a propagator amplitude (see, Refs.~\cite{BSh_NC_56}
and ~\cite{Book}). Here, $\gamma (g)$ is known as the {\it anomalous
dimension} of $S(x, g)$.
\par
Generally, the differential equation akin to (\ref{eq1.3})
\be x f_x - \beta(g) f_g = 0 \,.
\label{eq1.6} \ee
states an invariance of a function $f$ under the RG transformation
(\ref{eq1.1}).
Its solution $f(x,g)=F(G(x,g))$ precisely corresponds to the same property
emphasized by (\ref{eq1.2}).
\par
In a particular case, when the function $G$ is linear in the last argument,
$G \sim kg$, equation (\ref{eq1.2}) defines a solution that has a power $x$
dependence, i.e., $G(x,g)=gx^{k}$ with $k$ being an arbitrary number. Then,
equation (\ref{eq1.2}), takes a form of power scaling (or {\it power
self-similarity}) transformation
$$x'=x/a~, ~~~g'= g a^{k}~,$$
that is well-known in mathematical physics and widely used in the
problems of hydrodynamics of liquids and gases.
\par
Therefore, transformation (\ref{eq1.2}) can be considered \cite{Sh_TMF_84}
as a functional generalization $gx^k \to G(x, g)$ of a usual (i.e., power)
self-similarity transformation. One can refer to it as to {\it functional
self-similarity transformation}: this term was first introduced in
\cite{Sh_SPD_82} as a synonym of the RG transformation as defined above.
\par
It is widely known that in QFT, as well as in other mentioned fields
of theoretical physics, the RG method allows one to improve the perturbation
theory results and to simplify the analysis of a singular behavior of a
solution which becomes scale-invariant in the vicinity of a singularity.
The latter reminds a situation, which is typical of asymptotic analysis of
solutions of differential equations (DEs): long-time asymptotics demonstrate
self-similar regimes \cite{Baren79}. Hence, it looks natural to use the RG
methods to study strong nonlinear regimes and to investigate asymptotic
behavior of physical systems described by DEs. We have no possibility to
discuss this in detail here and would limit ourselves to mentioning some
successful attempts of using the RG ideas in mathematical physics.
\par
To our knowledge, the very first results in this field were obtained about a
decade ago by two of the co-authors~\cite{KP_PF78_87}
of this paper by applying RG ideas to a problem of generating of higher
harmonics in plasma.  This problem, after some simplification, was reduced to
a couple of partial DEs with the boundary parameter -- solution
"characteristic" -- explicitly included. It was proved that these DEs admit an
exact symmetry group similar to that, defined by Eq.(\ref{eq1.3}). The group
obtained was then utilized to construct the desired nonlinear solution of the
BVP. This approach has further been developed (see \cite{KP_TMP_90} and
references therein) and we shall describe it in some detail in the next
Section.
\par
The methods of QFT RG were exploited by Goldenfeld, Martin and Oono with
co-authors (Urbana group) ( see, e.g., Refs.~\cite{GMO_JSC_89} and
\cite{GMOL_PRL_90}) to find asymptotics of the solutions of parabolic-type
nonlinear differential equations, that describe a variety of physical
phenomena, such as groundwater flow under gravity, shock waves dynamics,
radiative heat transfer and so on.  As an auxiliary tool, they used the
concept of {\it intermediate asymptotics}\/ first introduced by Barenblatt
and Zeldovich~\cite{BZ_RMS_71} -- see also the review
monographs~\cite{Baren79} and \cite{Baren96}.  In this way, the Urbana team
was able to determine values of exponents in the ratios of invariants
forming arguments of self-similar solutions.
\par
Later on, with the goal to a global asymptotic analysis they developed and
illustrated, by numerous examples, the "perturbative renormalization group
theory" (see \cite{CGO_PhysRev_96} and references therein) that exploited the
form of the invariance condition akin to that (\ref{eq1.3'}) used in QFT. The
geometrical formulation of the perturbative RG theory for global analysis was
presented by Kunihiro \cite{Kunih} on the basis of a classical theory of
envelopes.
\par
On the other hand, Bricmont and Kupiainen~\cite{Bric_CMP_92} --
\cite{Bric_DE_96} attracted RG ideas for nonlinear DEs analyzing in a bit
different manner. They used an iterative set of rescalings borrowed from
the Wilson version of renormalization group, that is semi-group. On basis of
that RG-mapping procedure they succeeded in proving the global existence and
detailed long time asymptotics for classes of nonlinear parabolic equations.
\par
Generally, the procedure of revealing RG transformations, or some group
features, similar to RG regularities, in any partial case (QFT, spin lattice,
polymers, turbulence and so on) up to now is not a regular one. In practice,
it needs some imagination and atypical manipulation (see discussion in
\cite{Sh_IJMP_88,Sh_RMS_94,GMO_JSC_89}) "invented" for every particular case.
For example, the above described RG methods applied to asymptotic analysis of
differential equations were based on the  a priori assumption of the
existence of some scaling transformations or on the invariance condition of
an approximate solution.  By this reason, the possibility to find a regular
approach to constructing RG-symmetries is of principal interest. In this paper
we give an account of our efforts for creating a possible scheme of this kind
in application to physical systems that are described by DEs. The leading
idea in this case is based on the fact that symmetries
of such systems can be found in a regular manner by using the well-developed
methods of modern group analysis.
\par
The paper is organized as follows: in Section II we describe the
general scheme of constructing RG-symmetry for a BVP. It appears, that the
implementation of this scheme strongly depends on the mathematical model used
and on the form of boundary conditions. As a result, different approaches to
finding RG-symmetries are possible, and these are illustrated by examples in
the following five sections. In Section III RG-symmetries are calculated
using the classical Lie symmetries. In Section IV RG-symmetries are obtained
on the basis of Lie-B\"acklund symmetries. In the next two sections
RG-symmetries are found when boundary conditions are presented either in the
form of a differential constraint (Section V), or in the form of an
embedding equation (Section VI). In Section VII one more approach to
RG-symmetries constructing is presented which is based on an approximate
group symmetry.  In conclusion, we make a summary of the approach and
discuss some further applications.

\section{Approach to constructing RG-symmetries}
First of all, we emphasize that the desired regular approach to
constructing RG-symmetries turns out to be possible for those mathematical
models of physical systems that are based on differential or, in some
particular cases, integro-differential equations. The key idea uses the fact
\cite{KPS_PD447_95,Shr_Pisa-95} that such models can be analyzed by algorithms
of modern group analysis.
\par The proposed scheme comprises a sequence of the
four steps.
\par
{\bf I.} A specific manifold (differential, integro-differential, etc.)
should be primarily constructed.
This manifold that will be referred to as {\it renormgroup manifold}
(RG-manifold) generally differs (see below) from the manifold given by the
original system of DEs.
\par
{\bf II.} The second step consists in calculating the most general
symmetry group $\cal{G}$ admitted by the RG-manifold.
\par
{\bf III.}
The restriction of the group $\cal{G}$ on the desired BVP solution (exact
or approximate) constitutes the next step. The group of
transformations thus obtained (renormgroup) is characterized by a
set of infinitesimal operators (RG-operators), each containing the
solution of a BVP in its invariant manifold.
\par
{\bf IV.} The last, fourth step implies utilization of RG-operators to find
analytical expressions for solutions of the BVP.
\par
\vspace{1mm}
\par
Being formulated in a concise form these steps deserve further
comments.
\par
{\it Comment to I}.\/ In the scheme described above the first step,
namely constructing the RG-manifold, is of fundamental importance.
The form of its realization depends both on a mathematical model
and on a form of a boundary condition. \par
Here we show the following different approaches to RG-manifold
constructing:
\par
{\bf Ia.}
In the first, more simple case the RG-manifold, as usual in classical
group analysis, is presented by a system of basic DEs with the only
substantial difference: parameters, entering into a solution via the
equation and boundary conditions, are included in the list of independent
variables. \par
{\bf Ib.}
Another approach to constructing the RG-manifold implies an
extension of a space of variables involved in group transformations,
for example, by including differential variables of higher order
and nonlocal variables. It means that in this case Lie-B\"acklund
transformation groups and nonlocal transformation groups should be
invoked \cite{Ibr,Sprav}.
\par
{\bf Ic.}
In the third approach the procedure of construction of RG-manifold
is based on the invariant embedding method \cite{Ambar}.
Here RG-manifold is given by a system of equations that consists of
original DE and/or embedding equations which correspond to the BVP under
consideration.
\par
{\bf Id.}
The fourth approach to some extent is similar to the previous one.
In this event boundary conditions are reformulated in terms of a
differential constraint which is then combined with original equations
to form the desired RG-manifold.
\par
{\bf Ie.}
The last approach utilizes approximate transformation groups.
Here, the RG-manifold is given by a system of DEs with small
parameters and can be analyzed by perturbation methods \cite{BGI}.

\vspace{1mm}

{\it Comment to II}.\/ Searching the symmetry of RG-manifold is the main
problem of the second step. The term "symmetry" as used in the classical
group analysis means the property of a system of DEs to admit a Lie group of
point transformations in the basic space of all independent and dependent
(differential) variables entering these DEs.  The Lie calculational algorithm
of finding such symmetries is reduced to constructing tangent vector fields
with coordinates, that are functions of these basic group variables and
can be defined from the solution of an overdetermined system of DEs, named
as {\it determining equations}. In modern group analysis different
modification of a classical Lie scheme are in use (see,
e.g.\cite{Sprav,Olver} and references therein). If the problem of finding
symmetries for a given system of DEs (RG-manifold) is solved, then the
result is presented in the form of Lie algebra of infinitesimal operators
(also known as group generators), which correspond to the admitted vector
field. In what follows these operators will be denoted by $X$.

\vspace{1mm}

{\it Comment to III}.\/
The goal of a group restriction is the construction of a transformation
group with a tangent vector field (point, Lie-B\"acklund, etc.)
infinitesimal operators of which (hereinafter referred as $R$) contain the
desired BVP solution in an invariant manifold. This means, that the coordinate
of the canonical operator of RG-symmetry vanish on the BVP solution and on its
differential consequences.
\par
Mathematically, the procedure of a group restriction appears as a "combining"
of different coordinates of group generators $X$ admitted by the
RG-manifold.  The vanishing condition for this combination
on a solution of the BVP leads to algebraic equalities that
couple different coordinates and give rise to desired RG-symmetries. In a
particular case, when RG is constructed from a Lie group admitted by the
original system of DEs, it turns out to be a  subgroup of this group and a
solution of the BVP appears as an invariant solution with respect to the
point RG obtained (compare with \cite{Ovs82}). In the general case, not only
 Lie point group, but Lie-B\"acklund groups, approximate groups, nonlocal
transformation groups, etc. (see, e.g. \cite{Sprav}), are also employed as
basic groups which are then to be restricted on the solution of a BVP.

\vspace{1mm}
\par
{\it Comment to IV}.\/ A technique for constructing group invariant
solutions corresponding to a symmetry group when its infinitesimal operators
are known has been detailed in various monographs (see, e.g.,
\cite{Ovs82,Olver,Sprav}). Therefore, the final step is performed in a
usual way and needs no specific comments.

\vspace{3mm}

\par
Before proceeding any further, we make a short review of results, that were
obtained on the basis of the formulated scheme. The first application of
RG-approach to a particular problem of laser plasma was announced in
\cite{KP_PF78_87}. This problem, namely
the problem of a nonlinear interaction of a powerful laser radiation with
inhomogeneous plasma, has been  detailed in subsequent publications
\cite{KP_TMP_90,KP_KSF_89,KP_KSF_91}. A mathematical model was given by a
system of nonlinear DEs for components of electron velocity, electron density
and the electric and magnetic fields.  The presence of small parameters
(such as weak inhomogeneity of the ion density, low electron thermal
pressure and small angles of incidence of a laser beam on plasma surface) in
the initial system of equations provided a way to constructing RG-manifold
using ({\bf Ie}) approach, based on approximate group methods.
The desired RG-symmetry appears as Lie point symmetry that takes account
of transformations of a boundary parameter (common to ({\bf Ia}) approach),
which is related to the amplitude of the magnetic field at a critical
density point. RG-symmetry obtained made it possible to get the exact
solution of original equations, that was then used to evaluate the
efficiency of harmonics generation in cold and hot plasma (see
\cite{KP_TMP_90}).
\par
The advantageous use of the RG-approach in solving the above particular
problem gave promise that it may work in other cases. This was illustrated in
\cite{KKP_RG-91} by a series of examples for different BVPs. Various methods
of constructing RG-symmetries were described, based on the use of point
symmetries, approximate symmetries, embedding equations and transformations
of Fourier components. As is shown in \cite{KKP_RG-91},
different formulations of BVPs give rise
to various methods of finding RG-symmetries. Thus, further development of
the scheme was concentrated on analyzing these methods.
\par
The first one was concerned with the initial value problem for the modified
Burgers equation with parameters of nonlinearity and dissipation included
explicitly. This example \cite{KP_LGA_94} yielded a
detailed illustration of the method of constructing RG-symmetries when
a basic RG-manifold is given by an original DE with parameters included in
the list of independent variables ({\bf Ia} approach). It was argued that
the exact solution can be reconstructed from the perturbative solution with
the help of any of the admitted RG-symmetry operators which form an
eight-dimensional algebra. Two illustrative examples were given, dealing
with perturbation theory in time and in nonlinearity parameter.
\par
To demonstrate the method of constructing Lie-B\"acklund
RG-symmetries that uses ({\bf Ib}) approach,
the initial value problem for a linear parabolic equation was
considered in \cite{KPS_PD447_95}. It was shown that appending Lie-B\"acklund
RG-symmetries to point RG-symmetries extends the algebra of RG-symmetries up
to an arbitrary order.
\par
The same mathematical model was also employed within the
({\bf Ic}) approach when the boundary condition is described by a
differential constraint \cite{KKP_PF13_95}. It was found that RG-symmetries
obtained can not be reduced to point RG-symmetries which arise from the
({\bf Ia}) case.  However, some of them can be reformulated in terms of
RG-symmetries previously found in ({\bf Ia}) approach while the others
can be constructed from the Lie-B\"acklund symmetries of basic equations
in view of the given differential constraint.
\par
An idea of the ({\bf Id}) approach to constructing RG-manifold based on
the invariant embedding method was realized in \cite{KKP_RG-91} for ordinary
DEs. Here, embedding equations can be treated as a specific form of
a differential constraint, that takes boundary data into account. The
method of finding RG-symmetry using the ({\bf  Id}) approach
proves to be of particular interest for the first order ordinary
DEs when using ({\bf Ia}) approach faces standard problems in calculating
point symmetries admitted by RG-manifold. Provided the embedding equation
has the form of a first order DE of an evolutionary type there  appear no
difficulties in group analysis of a joint system of the basic and the
embedding equations.
\par
Worthy of mention is an example that demonstrates the utilization of
RG-symmetries to constructing solutions of the BVP for a system of two
first-order partial DEs that describes the propagation of a laser beam in a
nonlinear focusing medium
\cite{KPS_DE_93}--\cite{KS_PD41_97}.
It was revealed that RG-symmetries are related to formal symmetries
that are constructed in the form of infinite series in medium nonlinearity
parameter. For a specific form of boundary data infinite series are
truncated with RG-symmetries presented by finite sums. Generally, for
arbitrary boundary data this is not the case and in that event a finite sum
describes approximate RG-symmetry for small nonlinearity parameter. Based
on ({\bf Ia}), ({\bf Ib}) and ({\bf Ie}) approaches both point and
Lie-B\"acklund (exact and approximate) RG-symmetries were obtained and then
used to find an analytical solution of the problem.  \vspace{0.3cm}

\par
    To clarify the idea of constructing RG-symmetries several examples
are given below which demonstrate different approaches to the problem.
To gain better understanding of these approaches, a simple
mathematical model is used \cite{ftn4}. This model corresponds to BVP
for a system of two first-order partial DEs that were studied by Chaplygin
\cite{Chaplygin} in gas dynamics
\be
\begin{array}{c}
v_t+vv_x-a\varphi (n)n_x=0\,, \quad n_t+vn_x + nv_x = 0 \,; \\
\mbox{} \\
v(0,x)=V(x)\,, \quad 	  n(0,x)=N(x)\,, \end{array}
\label{eq2.1}
\ee
where $\varphi (n)$ is an arbitrary function of $n$ and $a$ is a
nonlinearity parameter. Despite its simplicity, this mathematical
model has a wide field of application and was used to describe
various physical phenomena (in the so-called quasi-gaseous
media \cite{ZhdTr_87}). In such a case the physical
meaning of variables $t$, $x$, $v$ and $n$ may differ from that in gas
dynamics.  For example, in nonlinear geometrical optics, $t$ and $x$ are
coordinates, respectively, along and transverse  to the direction of laser
beam propagation, $v$ is the derivative of eikonal  with respect to $x$, and
$n$ is a laser beam intensity. In this case, functions $V$ and $N$
characterize the curvature of the wave front and the beam intensity
distribution upon the coordinate $x$ and the entrance of a medium $t=0$.
\par
Along with (\ref{eq2.1}) another form of basic equations will be used
\be a \tau_v - (n/\varphi(n))\chi_n = 0\,, \quad
\chi_v + \tau_n = 0 \,.
\label{eq2.2}
\ee
These linear equations for new variables $\tau = nt$ and $\chi = x-vt$
results from (\ref{eq2.1}) under hodograph transformations.

\section{RG as Lie point subgroup}
This section presents an illustration of the method of constructing
RG-symme\-tries when a basic RG-manifold is given by the original DEs with
parameters included in the list of independent variables. Boundary  conditions
are taken into account while restricting the group admitted by RG-manifolds up
to the desired RG on the exact or approximate solution of a BVP which thus
appears as an invariant solution with respect to any of RG operators
obtained.
\par {\bf{3.1.}} First we shall consider a particular
case of equations (\ref{eq2.1}) when the nonlinearity parameter $a$ is equal
to zero: in application to optical equations discussed above this
means that nonlinear effects are neglected.  Then by introducing a
new variable $v=\varepsilon u$, the system of equations (\ref{eq2.1})
is rewritten in the following form
\be u_t+\varepsilon uu_x=0\,, \quad n_t+
\varepsilon un_x + \varepsilon nu_x = 0 \, ; \label{eq3.1} \ee \be
u(0,x)=U(x)\,, \quad n(0,x)=N(x)\,.
\label{eq3.1'} \ee
The continuous point Lie group admitted by the differential
manifold (\ref{eq3.1}) (RG-manifold) is given by the infinitesimal operator (a
general element of Lie algebra) with six independent terms
\be
X = \xi^1 \partial_t +\xi^2 \partial_x +\xi^3 \partial_{\varepsilon}
    +\eta^1 \partial_u +\eta^2 \partial_n \equiv \sum\limits_{i=1}^{6} X_i \, ,
\label{eq3.2} \ee
$$
\begin{array}{l} \displaystyle{
X_1= (1/\varepsilon)\, \Delta J^1 \, \partial_t + \left( J^1 + u\, \Delta J^1
\right) \partial_x -n J^1_{\chi} \partial_n \, , \quad
   \Delta J^k \equiv \left(\varepsilon tJ^k_{\chi} -J^k_u \right)\, , }\\
\mbox{} \\
\displaystyle{
X_2= (1/n)\, J^2 \left( \partial_t + \varepsilon
     u \partial_x \right) \,, \quad
X_3= n J^3 \partial_n  \,,  \quad
X_4 = J^4 \left( - t \partial_t
     + n \partial_n + \varepsilon \partial_{\varepsilon} \right) \, , } \\
\mbox{} \\
\displaystyle{ X_5= \Delta J^5 D + J^5 \left( \varepsilon t \partial_x +
\partial_{u}\right) \, ,  \quad
X_6= - (1/n) \,J^6 D\, , \quad D \equiv \left( t \partial_t
     + \varepsilon u t \partial_x - n \partial_n \right)   \, .  } \\
\end{array}
$$
Coordinates $\xi$ and $\eta$ of
this infinite-dimensional group operator depend upon five functions
$J^i(\chi,u,\varepsilon), \ i=1, 2, 3, 5, 6$ which appear as arbitrary
functions of their arguments $\chi=x-vt$, $u$ and $\varepsilon$. The
sixth one, $J^4$, that enters the operator which describes group
transformation of parameter $\varepsilon$, is an arbitrary function of
this parameter only. The restriction of the group admitted by
RG-manifold (\ref{eq3.1}) on the solution of the BVP
$u=\bar u(t,x,\varepsilon)$, $n=\bar n(t,x,\varepsilon)$ leads to zero
equalities for two coordinates of the operator (\ref{eq3.2}) in
the canonical form -- the conditions of functional self-similarity:
\be
\eta^1 +\xi^1 \varepsilon \bar u \bar u_x -\xi^2 \bar u_x -\xi^3
\bar u_{\varepsilon} = 0\,,  \quad
\eta^2 +\xi^1 \varepsilon (\bar n \bar u)_x -\xi^2 \bar n_x -\xi^3
\bar n_{\varepsilon} = 0\,.
\label{eq3.3} \ee
These equalities should be valid for any values of $t$, and
certainly for $t=0$, when dependencies $\bar u$ and $\bar n$ upon $x$ are
given by boundary conditions (\ref{eq3.1'}). This yields two linear
relations between $J^i$ and $J^1_{\chi}$:
\be   J^5= U_x J^1\, , \quad
J^6=  N_x J^1 + N J^1_{\chi} - N (U_x)_u J^1 - \varepsilon U_x J^2
- N J^3 - N J^4 \, .
\label{eq3.4} \ee
   Here, and in what follows functions $U$ and $N$ and their derivatives
with respect to $x$ should be expressed either in terms of $u$ or
in terms of $\chi$. Substituting (\ref{eq3.4}) in (\ref{eq3.2}) gives the
desired RG-symmetries with the RG-operator
\be R =   \sum\limits_{i=1}^{4} R_i  \, ,
\label{eq3.5} \ee
$$
\begin{array}{l}
\displaystyle{R_1=X_1+
\left[ \left( \varepsilon t ( U_x)_{\chi} - \left(1- \frac{N}{n} \right)
(U_x)_{u} - \frac{ N_x}{n} \right) J^1 \right. } \\ \mbox{} \\
\displaystyle{ \left.
\hphantom{R_1=} + \left(\varepsilon t U_x - \frac{N}{n}  \right)
J^{1}_{\chi} - U_x J^{1}_{u} \right] D +
U_x J^1 \left(\varepsilon t\partial_x +\partial_u \right)
\, ,
} \\
 \mbox{} \\
\displaystyle{
R_2= X_2 + \frac{\varepsilon U_x}{n} J^2 D ,  \quad
R_k= X_k + \frac{N}{n}J^k D \, , \quad k = 3,4 \,.
}
\end{array}
$$
We see  that RG-symmetries for (\ref{eq3.1}), (\ref{eq3.1'})
are presented as a combination of symmetries of infinite-dimensional
algebra with the infinitesimal operator (\ref{eq3.2}). Any of the four
operators $R_k$ (and their linear combinations with coefficients
that are arbitrary functions of $\varepsilon$) contains the BVP solution
$u=\bar u(t,x,\varepsilon)$ and $n=\bar n(t,x,\varepsilon)$
in the invariant manifold and enables to obtain group transformation
of both group variables and different functionals of the solution
(for a method of calculating transformation of a functional see Ref.
\cite{KKP_JNMP_96}).
\par
Generally, the renormalization group using is capable of improving
a perturbation theory solution.  As an example, consider the
perturbative solution of (\ref{eq3.1}), (\ref{eq3.1'}) for
small value of $\varepsilon t \ll 1$
\be
u = U(x) -(\varepsilon t) UU_x + O\left( \varepsilon^2 t^2 \right) \,,\quad
    n = N(x) - (\varepsilon t)\left( U N_{x} + N U_x \right)
       + O \left( \varepsilon^2 t^2 \right) \,.
\label{eq3.6} \ee
This approximate solution  in the limit $(\varepsilon t )
\rightarrow 0$ is invariant with respect to RG transformation defined
by the operator $R_2$ with arbitrary $\varepsilon J^2 \neq 0$.  Assuming
$J^2 =1/ \varepsilon$, we obtain the explicit expression for RG-operator
\be
R= \frac{1}{n} \left[ (1+\varepsilon t U_x )\left (\partial_t
   + \varepsilon u \partial_x \right) - \varepsilon U_x n \partial_n \right]  \,,
\label{eq3.7} \ee
and invariance conditions written in the form of two first order DEs:
\be
   u_t + \varepsilon u u_x =0 \,, \qquad
   (1+\varepsilon t U_x )(n_t + \varepsilon u n_x) + \varepsilon n U_x = 0 \,.
\label{eq3.8} \ee
Solving Lie equations  which correspond to RG-operator (\ref{eq3.7})
(and coincide with characteristics equations for (\ref{eq3.8}))
enables to
reconstruct the desired exact solution of (\ref{eq3.1}),
(\ref{eq3.1'}) from the perturbative solution (\ref{eq3.6})
\be
u=U(x-\varepsilon u t)\,, \quad
n=\frac{1}{1+\varepsilon t U_x} N(x-\varepsilon t u) \,,
\label{eq3.9} \ee
where $U_x$ should be expressed in terms of $u$.
For example, in particular case of $N(x)=N_0 \exp (-x^2)$, $U(x)=-x$ and
$\varepsilon=1/T$ the latter formulas describe the focusing of gaussian
laser beam in geometrical optics
\be n = \frac{T}{T-t} N_0\exp \left(
        -x^2\left(\frac{T}{t-T}\right)^2 \right) \,, \quad
    u = x\frac{T}{t-T} \,, \quad  t \leq T .
\label{eq3.10} \ee
\medskip
\par
{\bf 3.2.} Now let us turn to a more general case of $a \neq 0$.
The Lie point symmetry group, admitted by RG-manifold
(\ref{eq2.2}), is characterized by a canonical infinitesimal operator
\cite{Ibr} with six independent terms $X_i,\ i=1,\ldots , 5$~and~$X_{\infty}$
\be
X=X_{\infty}+\sum\limits_{i=1}^{5}c_iX_i  \equiv
  \left( \bar f +\sum\limits_{i=1}^{5}c_if_i \right) \partial_{\tau} +
  \left( \bar g +\sum\limits_{i=1}^{5}c_ig_i \right) \partial_{\chi} ,
\label{eq3.12}
\ee
where coordinates $f_i$ and $g_i$ are linear combinations of
$\tau$ and $\chi$ and their first derivatives  $\tau_1 = (\partial \tau /
\partial n ) $ and $\chi_1 = ( \partial \chi / \partial n )$ with
coefficients depending only on $v$ and $n$ \cite{K_JNMP_96,KP_MCM_97}.
For a particular case $\varphi = 1$ they are
\be
\begin{array}{l}
f_1=\tau, \quad g_1=\chi; \quad f_2= - (1/a)n\chi_1, \quad g_2=\tau_1;\\
\mbox{} \\
\displaystyle{
f_3=-\tau /2 + n \tau_1 + (1/2a)nv\chi_1, \quad
g_3=-(v/2)\tau_1 + n\chi_1;  }\\
\mbox{} \\
\displaystyle{
f_4= -(1/2)n\chi+vn\tau_1 + \left[(1/4a)v^2 - n\right]n\chi_1,}\\
\mbox{} \\
\displaystyle{
g_4= (a/2)\tau + (1/2)v\chi
     +vn\chi_1
 +\left[ an-(1/4)v^2 \right]\tau_1; } \\
\mbox{} \\
\displaystyle{ f_5=\left(n\tau_1-\tau\right) -a\tau_a \, , \quad
               g_5=n\chi_1 -a\chi_a \,.}
\end{array}
\label{eq3.13} \ee
"Evident" symmetries $f_1,\ g_1$ and $f_2, \ g_2$ describe dilations of
$\tau$ and $\chi$ and translations along $v$-axis respectively for an
arbitrary nonlinearity $\varphi(n)$.  Two more symmetries $f_3,\ g_3$ and
$f_4, \ g_4$ appear due to a special form of the function $\varphi =1$ under
consideration. The symmetry $f_5,\ g_5$ involves the parameter $a$
transformation along with transformations of dynamic variables.
\par
The operator $X_{\infty}$ with coordinates $\bar f = \xi^{1}(v,n), \
\bar g =\xi^{2}(v,n) $
that are arbitrary  solutions of partial DEs
\be
\xi^{1}_v - (n/a)\xi^{2}_n = 0 , \quad \xi^{2}_v + \xi^{1}_n = 0 \,,
\label{eq3.14}
\ee
results from the linearity of basic Eqs.(\ref{eq2.2}); it is an ideal
of an infinite-dimensional Lie algebra $L_{\infty}$ formed by
operators $X_1, \ldots , X_5$ and $X_{\infty}$.
\par
The restriction of the group (\ref{eq3.13}) on the BVP solution means
that coordinates $f$ and $g$ of the canonical operator
(\ref{eq3.12}) turns to zero on this solution, that is
\be  \bar f = - \sum\limits_{i=1}^{5}c_if_i , \quad
     \bar g = - \sum\limits_{i=1}^{5}c_ig_i \, .
\label{eq3.15} \ee
These relations express functions $\bar f $ and $\bar g$ in
terms of $f_i,\ g_i, \ i=1,\ldots , 5$ taken on a solution $\tau =
\bar{\tau} (v,n)$, $\chi = \bar{\chi} (v,n)$ of a BVP (exact or
approximate).
Substitution of (\ref{eq3.15}) in (\ref{eq3.12}) gives five RG-operators
\be
R=\sum\limits_{i=1}^{5}c_i(a)R_i \, ,
\label{eq3.16} \ee
each being determined by corresponding coordinates $f_i, \ g_i$
and by a pair of functions $A^i,\ B^i$
\be
\begin{array}{l}
\displaystyle{
R_1= (\tau-A^1) \partial_{\textstyle{\tau}}
    +(\chi-B^1) \partial_{\textstyle{\chi}} \,, \quad
R_2= - A^2 \partial_{\textstyle{\tau}} -B^2  \partial_{\textstyle{\chi}}
     + \partial_{\textstyle{v}} }\,,\\
    \mbox{} \\
    \displaystyle{
R_3= (-(\tau/2)-A^3) \partial_{\textstyle{\tau}}
    -B^3 \partial_{\textstyle{\chi}} -(v/2) \partial_{\textstyle{v}}
    -n \partial_{\textstyle{n}} }\,,\\
    \mbox{} \\
    \displaystyle{
R_4= \left(-(n/2)\chi -A^4\right) \partial_{\textstyle{\tau}}
    +\left((a/2)\tau +(v/2)\chi-B^4\right) \partial_{\textstyle{\chi}}}\\
    \mbox{} \\
    \displaystyle{
\hphantom{R_3= }
    +\left(-(1/4)v^2 + an \right) \partial_{\textstyle{v}}
    +vn \partial_{\textstyle{n}} }\,,\\
     \mbox{} \\
     \displaystyle{
R_5= \left( -\tau-A^5 \right) \partial_{\textstyle{\tau}}
     -B^5 \partial_{\textstyle{\chi}}
     -n \partial_{\textstyle{n}} +a \partial_{\textstyle{a}} \,}
\end{array}
\label{eq3.17} \ee
Here, ten functions $A^i, B^i$ are defined by expressions (\ref{eq3.13})
for $f^i$ and $g^i$ where one should replace $\tau$, $\chi$
by $\bar \tau (n,v)$, $\bar \chi (n,v)$.
Explicit formulas for RG-operators depend upon the specific solution of
the BVP. For example, for the particular solution of the BVP
(\ref{eq2.1}) with $V=0$ and $N(x)=\cosh^{-2}(x)$, described by
\cite{Akhmanov}
$$ \tau =
\frac{(v/2)^{1/2}}{a^{3/4}}{\left(\sqrt{\kappa^2+1} -\kappa\right)}^{1/2}
\,, \qquad \kappa = \frac{\sqrt{a}}{v}\left( 1  - n - \frac{v^2}{4a}\right),
$$
\be \!\!\! \chi=-\frac{1}{2}\ln \frac {(v/2\sqrt{a})^{1/2} +
{\left(\sqrt{\kappa^2+1}-\kappa \right)}^{1/2}} {-(v/2\sqrt{a})^{1/2} +
{\left(\sqrt{\kappa^2+1}-\kappa \right)}^{1/2}}\,,  \label{eq3.18}
\ee
functions $A^5,\ B^5$ in (\ref{eq3.17}) are expressed as follows
\cite{K_JNMP_96}:
$$
\begin{array}{l}
{\displaystyle
A^5=-\frac{{(v/2)}^{1/2}}{4a^{3/4}\sqrt{1+\kappa^2}}
         {\left(\sqrt{\kappa^2+1} -\kappa\right)}^{1/2}
\left(\kappa+\sqrt{1+\kappa^2}-
         2\frac{\sqrt{a}}{v}\right)\,,} \\
\mbox{} \\
{\displaystyle
B^5=\frac{{(v/2)}^{1/2}}{4a^{1/4}\sqrt{1+\kappa^2}}
\frac
{ {\left(\sqrt{\kappa^2+1}-\kappa \right)}^{1/2}}
{( \sqrt{\kappa^2+1}-\kappa  -(v/2\sqrt{a})) }
\left(\sqrt{1+\kappa^2} - 3\kappa + 2\frac{\sqrt{a}}{v}
 - \frac{v}{\sqrt{a}} \right)\,.}
\end{array}
$$
\medskip
\par
It should be noticed, that the solution of the presented above BVP is
unique, but a number of RG-operators that give rise to this solution is
different from one (in the first example case we have four RG-operators
with arbitrary functions of $(n, \ \chi)$, and in the second example five
RG-operators with arbitrary functions of $a$).
In the next section we will show
that the number of RG operators may be enlarged to an arbitrary value,
provided not only point but Lie-B\"acklund groups are taken into account.

\section{RG as Lie-B\"acklund subgroup}
The method of constructing RG-symmetries from Lie point symmetries
admitted by the original DE is naturally generalized to include
Lie-B\"acklund (L-B) symmetries. The extension of the space of differential
variables increases the amount of BVPs that allow
restriction of a group on their solution. A complete set of RG-symmetries
is obtained by appending L-B RG-symmetries to point
RG-symmetries. In this section we present an example of
constructing L-B RG-symmetries of the second order for
the BVP (\ref{eq2.1}). As in the previous section we use
a transformed form of the basic equations (\ref{eq2.2}).
\par
L-B symmetries admitted by the RG-manifold (\ref{eq2.2})
are characterized by the same canonical infinitesimal operator
(\ref{eq3.12}) where additional terms
proportional to higher-order derivatives of $\tau$ and $\chi$ should be
added in coordinates $f$ and $g$.  Similarly to first-order symmetries,
these terms are linear combinations of $\tau$ and $\chi$ and their
derivatives $\tau_i = (\partial^i \tau / \partial n^i)$  and
$\chi_i = (\partial^i \chi / \partial n^i) $ with coefficients that depend
only on $v$ and $n$ \cite{K_Pisa-95,K_JNMP_96,KP_MCM_97}.  For the
second-order Lie-B\"acklund symmetries in a particular case $\varphi(n)=1$,
we have five additional operators $X_i$ with $i=7, \ldots , 11$ (the term
with $i=6$ corresponds to $X_{\infty}$ and is omitted in the sum, i.e. $c_6
=0$)
\be \label{eq4.1} X=X_{\infty}+\sum\limits_{i=1}^{11}c_iX_i  \equiv
\left( \bar f +\sum\limits_{i=1}^{11}c_if_i \right) \partial_{\tau} +
\left( \bar g +\sum\limits_{i=1}^{11}c_ig_i \right) \partial_{\chi}\,.
\ee
It should be noted that expressions for all coordinates in (\ref{eq4.1})
can be obtained by the  action  of the following three recursive operators
\cite{KP_MCM_97} $L_i, \ i=1,2,3$
$$
  L_1= \left( \begin{array}{cc} 0 & -(n/a) D_{n} \\
\mbox{}\\ D_{n}   &      0 \end{array} \right) , \quad
  L_2= \left( \begin{array}{cc} \displaystyle{
2n D_n -1 } & \displaystyle{ (n/a) vD_{n}}\\
\mbox{}\\
\displaystyle{ - vD_{n}}  & \displaystyle{ 2 n D_n }
\end{array}
\right) \,,
$$
\be
L_3=
\left(
\begin{array}{cc}
  \displaystyle{2 n vD_n }
&  \\
\mbox{}\\
\displaystyle{(-v^2/2 + 2 an ) D_{n} + a } &
\end{array}
\begin{array}{cc}
& \displaystyle{
n(v^2/2a - 2n ) D_{n} -n } \\
\mbox{}\\
& \displaystyle{
2n vD_n + v}
\end{array}
\right) ,
\label{eq4.2}
\ee
on the "trivial"  operator with $f=\tau$ and $g=\chi$ (here, $D_n$ is the
operator of total differentiation with respect to $n$).
Below, we present only
three of these five second-order L-B operators
\be
\begin{array}{l} \displaystyle{
f_7=n\tau_2, \qquad
g_7=\chi_1 + n\chi_2 ;}\\
\mbox{} \\
\displaystyle{
f_8=(1/2a)n\left[-\chi_1+v\tau_2 -2n\chi_2\right], \qquad
g_8= (1/2a)v\chi_1 +n\tau_2 + \frac{1}{2a}nv\chi_2 ; } \\
\mbox{} \\
\displaystyle{
f_9=(1/4)\tau-n\tau_1-(5/4a)vn\chi_1
    +\left(-n+(1/4a)v^2 \right)n\tau_2
     - (1/a)vn^{2}\chi_2, } \\
\mbox{} \\
\displaystyle{
g_9=(3/4)v\tau_1
     -\left(2n
 -(1/4a)v^2\right)\chi_1
+ vn\tau_2 +  \left(-n + (1/4a)v^2\right)n\chi_2 \,.}
\end{array}
 \label{eq4.3}
\ee
The procedure of restriction of the L-B group obtained on the
solution of the BVP leads to expressions for $\bar f$ and
$\bar g$ akin to (\ref{eq3.15})
\be
\bar f = - \sum\limits_{i=1}^{11}c_if_i , \quad
\bar g = - \sum\limits_{i=1}^{11}c_ig_i \, .
\label{eq4.4}
\ee
Substitution of (\ref{eq4.4}) in (\ref{eq4.1}) yields additional terms in the
expression (\ref{eq3.16}) for the RG-operator $R$ that depends on
higher-order derivatives of $\tau$ and $\chi$
\be
R=\sum\limits_{i=1}^{11}c_i(a)R_i \equiv
  \sum\limits_{i=1}^{11}c_i(a) \left(
   (f_i-A^i) \partial_{\textstyle{\tau}}
   + (g_i-B^i) \partial_{\textstyle{\chi}} \right) \, .
\label{eq4.5}
\ee
Here functions $A^i$ and $B^i$ are given by the corresponding formulas  for
coordinates $f_i$ and $g_i$ to be evaluated on the solution
$\bar \tau (n,v)$ and $\bar \chi (n,v)$. It appears that coordinates of
L-B RG-operators are obtained from point RG-operators with the help
of the above-mentioned recursive operators, hence, one can obtain
L-B RG-operators of an arbitrary high order. Despite an unusual
form, we still call them RG-operators since they possess the main property of
RG-operators, namely, they contain a solution of the BVP in their
invariant manifold.
\par
The procedure of using L-B RG-operators is not as simple
as for point RG-operators. Yet we can describe two possible ways.
\par
Firstly, coordinates of canonical L-B RG-operators can be used to construct
a set of relations, differential constraints, that are compatible with the
original DEs and satisfy specific boundary conditions.  The use of such
constraints is described in section 6.  In the general case, for an
arbitrary L-B group of a given order, coordinates of the corresponding
canonical operator can be treated as a set of differential expressions, zero
equalities for which impose appropriate restrictions on the basic DEs,
consistent either with physical or with symmetry conditions. These
equalities can also be treated as embedding equations (see
\cite{KPS_PD447_95}).
\par
Secondly, L-B RG-operators can be used to construct invariant solutions
that automatically fit boundary conditions.  It should be noticed that in
some particular cases, L-B RG-symmetries can be constructed from a L-B group
with a finite number of operators.  For example, RG-symmetry for
(\ref{eq2.1}) with boundary conditions $V=0$ and $N=\cosh^{-2}(x)$ appears
as a linear combination of three L-B symmetries
\be R=(f_3+2(f_7 + f_9)) \partial_{\tau}
   +(g_3+2(g_7 + g_9)) \partial_{\chi}\,.
\label{eq4.6} \ee
The desired solution of the BVP can be found as the invariant solution with
respect to RG-operator (\ref{eq4.6}) and is presented by formulas
(\ref{eq3.18}).
\par The recipe of constructing the L-B
renormgroup formulated in this section goes far beyond a simple
illustrative example for the BVP  (\ref{eq2.1}). In a similar way,
L-B RG-operators are constructed for
different BVPs of mathematical physics that admit L-B
symmetries; other examples are presented in \cite{KPS_PD447_95} for the
linear parabolic and modified Burgers equation.
It is essential that when parameters entering into the equation and boundary
conditions are involved in group transformations,
coordinates of canonical L-B RG-operators contain not only first
but higher-order derivatives with respect to these parameters.
This means that in addition to recursive operators containing operators of
total differentiation with respect to $n$ (for BVP (\ref{eq2.1})), new
recursive operators comprise operators of
differentiation with respect to parameters, as well ($\propto D_{\epsilon}$
and $D_a$ in the case of BVPs (\ref{eq3.1}) and (\ref{eq2.2})).

\section{RG devising based on embedding equations}
In this section we present a specific method of constructing
RG-symmetries \cite{KKP_RG-91,KPS_PD447_95}, which is based on embedding
equations \cite{Ambar}. It is of prime interest for physical systems
described by ordinary differential equations (ODEs). In the context of the
discussed model of quasi-Chaplygin media such equations arise, e.g., when
constructing invariant solutions with respect to symmetries obtained. We
demonstrate the idea of this method for the very simple BVP
\be u_t=f(t,u,a)\,; \qquad t=\tau\,, \quad u=x\,.
\label{eq5.1} \ee
Extension of the original differential manifold by adding, to the original
equation, the embedding equation that appears as a linear first-order
partial DE
\be
u_{\tau}+f(\tau ,x,a)u_x=0 \,,
\label{eq5.2}  \ee
gives the desired RG-manifold, where $u$ is now treated as the function of
four variables $\{t,\, \tau,\, x,\, a \}$.  Performing the group analysis
for this RG-manifold involves boundary data and parameter $a$ in group
transformations, while the subsequent restriction of the group obtained on
any solution of the BVP yields the desired RG-symmetries. We give two
examples of such calculations for $f=au^2$ and $f=u^2+au^3$.
\par
{\bf{5.1.}}
In the event of $f=au^2$ the RG-manifold (\ref{eq5.1})-(\ref{eq5.2}) is given
by two equations
\be u_t=a u^2 \,, \quad u_{\tau} + a x^2 u_x=0 \label{eq5.3} \ee
that admit an infinite-dimensional Lie point algebra with five
independent elements
\be
\begin{array}{c}
\displaystyle{
X= \sum\limits_{i=1}^{5}\alpha_i X_i\,, } \\
\mbox{} \\
\displaystyle{
X_1= \partial_t + au^2 \partial_u \,, \quad
X_2= \partial_{\tau} + ax^2 \partial_x \,, } \\
\mbox{} \\
\displaystyle{
X_3=  u^2 \partial_u \,, \quad
X_4= x^2 \partial_x \,, \quad
X_5=  x^2\tau\partial_x + u^2 t \partial_u +
     \partial_a  \,.}
\end{array}
\label{eq5.4} \ee
Here, functions $\alpha_1$ and $\alpha_2$ depend upon five  variables
 $\{ t,\, \tau,\, x, \,a,\,u\}$, whereas $\alpha_i$, $i=3,4,5$
are arbitrary functions of three combinations$at+(1/u)$,
$a\tau+ (1/x)$, $a$.
\par
The procedure of restriction of the group obtained leads to the invariance
condition
\be
U^2(\alpha_3+a\alpha_1+\alpha_5 t) -\alpha_1 U_t - \alpha_2 U_\tau
-x^2(\alpha_4+a\alpha_2+\alpha_5\tau)U_x -\alpha_5 U_a  = 0
\label{eq5.6} \ee
to be fulfilled on an exact or approximate solution
$u=U(t,x,\tau,a)$ of the BVP (\ref{eq5.1})-(\ref{eq5.2});
for example,  one can take the perturbative solution
as an expansion in powers of ~$a$
\be u = U(t,x,\tau,a) \equiv x+ax^2(t-\tau )+O(a^2)~, \qquad a \ll 1 \,.
\label{eq5.7} \ee
Substituting (\ref{eq5.7}) into (\ref{eq5.6}) shows that the invariance
condition (\ref{eq5.6})
is fulfilled for $\alpha_3 = \alpha_4 \equiv
\alpha$ and arbitrary $\alpha_1, \alpha_2$ and $\alpha_5$.  Assuming
 $\alpha_1=\alpha_2 = \alpha = 0$ and $\alpha_5=1$ in (\ref{eq5.4})
yields one of the RG-operators
\be R= x^2 \tau \partial_x + \partial_a + u^2 t \partial_u \,,
\label{eq5.8} \ee
which enables us to  transform the perturbative solution of
(\ref{eq5.1}) for small $a \ll 1$ to the following  exact solution
$$
u=\frac{x}{1-ax(t-\tau)}\,.
$$
This result is found by solving the Lie equations,  that correspond to the
RG-operator (\ref{eq5.8}).
\par {\bf 5.2.}
For another value of the function  $f=u^2+au^3$, the RG-operator that is
similar to (\ref{eq5.8}) is given as follows
\be
R= \left( x^2(1+ax)\tau +x \right) \partial_x
 + \left( u^2 (1+au)t +u \right)\partial_u - a\partial_a  \,.
\label{eq5.10} \ee
The invarince condition for the solution of the BVP
with respect to the RG-operator (\ref{eq5.10}) has the form of the first-order
partial DE
\be  - \left( x^2(1+ax)\tau +x \right) u_x + a u_a + u^2 (1+au)t +u  = 0 \,.
\label{eq5.11} \ee
Solving the characteristic equations for (\ref{eq5.11}) (Lie equations)
yields the following exact solution of the BVP (\ref{eq5.1}) with
$f=u^2+au^3$
$$
t-\tau = \frac{1}{x} -
\frac{1}{u} + a \ln \Big\vert \frac{x}{u}\frac{(1+au)}{(1+ax)} \Big\vert \,.
$$
\par
What all renormgroups obtained for the BVPs for the first-order ODE in the
above examples have in common is that their operators depend upon arbitrary
functions $\alpha_i$, which means that RG can be expressed in terms of
different RG-operators with various particular expressions for their
coordinates. This situation is the same as that one obtains for the BVP in
the case of partial DE:  different RG-operators yield the same
unique specific solution of the given BVP contained in the
invariant manifold of RG-operators. The previous procedure of RG
constructing for the BVP for the ODE was based on the use of point
groups. However, L-B groups can also be employed for constructing
RG-symmetries for the first-order ODE, especially, in view of embedding
equations (see Refs. in~\cite{KPS_PD447_95}).
\par
The structure of embedding equations depends not only on the form of the
original equation, but also on the boundary conditions. This means that
for given basic equations we may obtain different embedding equations.
For example, if the function
$f$ in the r.h.s. of (\ref{eq5.1}) depends upon $x$,
\be u_t=f(t,x,a,u)\, ; \quad t=\tau\,, \  u=x
\label{eq5.13} \ee
we arrive at the embedding equation
\be
\begin{array}{l}
\displaystyle{ u_{\tau} + f(\tau,x,a,x) u_x = f(\tau,x,a,x) }
\\ \mbox{}\\
\hphantom{u_{\tau + f(\tau,x,a,x) u_x}}
\displaystyle{ + \int\limits_{\tau}^{t} d\,t' f_x(t',x,a,u(t'))
\exp\left[ -\int\limits_{t}^{t'} d\,t'' f_u(t'',x,a,u(t'')) \right] \,.}
\end{array}
\label{eq5.14} \ee
Hence, the RG manifold in this case is  defined by a system of
integro-differen\-tial equations (\ref{eq5.13}) and (\ref{eq5.14}) and one
should employ the modern group analysis techniques which give a possibility
of analyzing such equations, as well~\cite{Mel,KKP_DE_93}.

\section{RG and differential constraint}
In the previous section RG-manifold was obtained by combining an original DE
and an embedding equation. More generally instead of an embedding
equation, an additional differential constraint can be used that satisfy two
conditions: firstly, it must be compatible with the original DE and,
secondly, it should explicitly take boundary conditions into account.
This constraint naturally emerges when a coordinate of a canonical operator
of the L-B RG admitted by BVP is assumed to be equal to zero. Adding
this constraint to original equations we obtain the RG-manifold.
\par
{\bf 6.1.}
To illustrate, consider first a BVP
(\ref{eq2.1}) with $a=0$ which we rewrite using hodograph transformations in
a simple form (compare with (\ref{eq2.2}))
\be \chi_n = 0\,, \quad \chi_v + \tau_n = 0 \,.
\label{eq6.1} \ee
L-B symmetries of this system of DEs are given by a canonical operator
\be X=f\partial_{\tau} + g\partial_{\chi}\,,
\label{eq6.2} \ee
with coordinates $f$ and $g$ depending upon $v$ and derivatives
$\tau_s+n\chi_{s+1}$, $\chi_s$ of an arbitrary order $s\geq 0$
\be
\begin{array}{c}
\displaystyle{ f = F(v,\chi_s,\tilde{\tau}_s)
              - n \left[ \partial_v + \sum\limits_{k=0}^{\infty}
             (\tilde{\tau}_{k+1} \partial_{\tilde{\tau}_s}
              +\chi_{k+1} \partial_{\chi_k}) \right] G , \quad
              g = G(v,\chi_s,\tilde{\tau}_s) \,,} \\
\mbox{} \\
\displaystyle{\tilde{\tau}_s = \tau_s + n \chi_{s+1} \, , \quad
              \tau_s = \left( \partial^s \tau / \partial v^s \right) \,,
\quad \chi_s = \left( \partial^s \chi / \partial v^s \right) \, .}
\end{array}
\label{eq6.3} \ee
Consider a particular case of a BVP (\ref{eq2.1}) with boundary conditions
defined by $V(x)=-\varepsilon x$ and arbitrary $N(x)$. In terms of the
variables $\tau$ and $\chi$, these conditions are described, for example, by
a pair of differential constraints
\be
\chi_{vv} = 0 \,, \quad  \tau_{vv} - N_{vv} \chi_{v} - N_v \chi_{vv} = 0\,.
\label{eq6.4} \ee
Here the dependence of $N$ upon $x$ is given in terms of $v$ with the use of
the above boundary condition.
\par
It is easily checked by direct substituting into (\ref{eq6.3}) that left-hand
sides of these equalities are the corresponding coordinates $g$ and $f$
of the second-order L-B symmetry operator (\ref{eq6.2}).
Adding differential constraints (\ref{eq6.4}) to the original equation
(\ref{eq6.1}), we obtain the desired RG-manifold
\be
\begin{array}{c}
\chi_n = 0\,, \quad \chi_v + \tau_n = 0 \,, \quad
\chi_{vv} = 0 \,, \quad  \tau_{vv} - N_{vv} \chi_{v} = 0\,.
\end{array}
\label{eq6.5}
\ee
The latter admits a 17-parameter group of point transformations given by the
following operators
\be
X=\sum\limits_{i=1}^{m} c_i X_i\,, \quad m=17 \,,
\label{eq6.6}
\ee
$$
\begin{array}{l}
\displaystyle{
X_{1}= v^2 \partial_{v} + v (2(n-N)+vN_v) \partial_{n}
   + (\chi(N-n) + \tau v) \partial_{\tau}
   + v \chi \partial_{\chi} \,, }\\
\mbox{}\\
\displaystyle{
X_{2}= v\chi \partial_{v} +  (\chi (n-N) + v (\chi N_v - \tau )) \partial_{n}
   + 2\tau \chi \partial_{\tau}
   + \chi^2 \partial_{\chi} \,, }\\
\mbox{}\\
\displaystyle{
X_{3}= -v \partial_{v} +  ( N-n - v N_v ) \partial_{n} \,, \quad
X_{4}= v\chi \partial_{n} - \chi^2 \partial_{\tau}\,, }\\
\mbox{}\\
\displaystyle{
X_{5}= v \partial_{n} \,, \quad
X_{6}= ( N-n ) \partial_{n} + \chi \partial_{\chi} \,, \quad
X_{7}= (n- N ) \partial_{n} + \tau \partial_{\tau} \,,}\\
\mbox{}\\
\displaystyle{
X_{8}= \partial_{\tau} \,, \quad
X_{9}= v \partial_{\tau} \,, \quad
X_{10}= (N-n) \partial_{\tau} + v \partial_{\chi} \,, } \\
\mbox{}\\
\displaystyle{
X_{11}= \partial_{\chi} \,, \quad
X_{12}= \chi \partial_{\tau} \,, \quad
X_{13}= - v^2 \partial_{n}  + v \chi \partial_{\tau} \,. } \\
\mbox{}\\
\displaystyle{
X_{14}= - \partial_{v} -  N_v \partial_{n} \,, \quad
X_{15}= \partial_{n} \,, \quad
X_{16}= \chi \partial_{n} \,, \quad
X_{17}= -\chi \partial_{v} +  ( \tau - \chi N_v ) \partial_{n} \,. } \\
\end{array}
$$
The usual procedure of restriction of the group obtained on a solution of
the BVP  (\ref{eq6.1}) relates different coefficients in the sum
(\ref{eq6.6}) and gives the desired RG operators
\be
R=\sum\limits_{i=1}^{13} c_i R_i\,,
\label{eq6.7}
\ee
$$
\begin{array}{l}
\displaystyle{
R_1= X_1 \,, \quad R_2=X_2 \,, \quad
R_3= X_3 +\varepsilon X_{17} \,, \quad
R_4= X_4  \,, } \\
\mbox{}\\
\displaystyle{
R_5= X_5 + \epsilon X_{16} \,, \quad
R_{6}= X_6 +\varepsilon X_{17}  \,, \quad
R_{7}= X_{7} +\varepsilon X_{16} \,,} \\
\mbox{}\\
\displaystyle{
R_{8}= X_{8} +\varepsilon X_{15} , \quad
R_{9}= X_{9} -\varepsilon^2 X_{16} , \quad
R_{10}= X_{10} -\varepsilon^2 X_{17}, } \\
\mbox{}\\
\displaystyle{
R_{11}= X_{11} +\varepsilon X_{14} \,, \quad
R_{12}=X_{12}  +\varepsilon X_{16}  \,, \quad
R_{13}= X_{13} \,. } \\
\end{array}
$$
The exact solution of the BVP $\chi=-v/\varepsilon$, $\tau=(1/\varepsilon)
(n-N)$ is found either by solving Lie equations corresponding to any of
these RG-operators, or as the intersection of all invariant manifolds.
\medskip
\par {\bf 6.2.} Now let us turn to a general case of a
BVP (\ref{eq2.1}) with $a \neq 0$.  We shall consider the problem of
constructing RG-symmetries using the RG-manifold given by basic equations in
the form (\ref{eq2.2}) and the most simple differential constraint yielded
by the linear combination of the second order L-B symmetry
(\ref{eq4.3}) $f_7,\, g_7$ and trivial infinite-dimensional symmetry
$f_{\infty}=0,\, g_{\infty}=-1$
\be a \tau_v - n \chi_n = 0\,, \quad \chi_v + \tau_n = 0 \,, \quad
    n\tau_{nn} = 0 \,, \quad \chi_{n} + n\chi_{nn} - 1 = 0 \,.
\label{eq6.8} \ee
This differential  constraint describes, in particular, a linear dependence of
$N$ upon $x$ and $V(x)=0$.  The Lie point group admitted by the RG-manifold
(\ref{eq6.8}) is characterized by seven infinitesimal operators (use
the formula (\ref{eq6.6}) for $m=7$)
$$ \begin{array}{l}
\displaystyle{ X_1= -2v \partial_{v} - 4n \partial_{n} - 6 n (v/a)
\partial_{\tau} + (2 \chi - 6n +3v^2/a) \partial_{\chi} \,, } \\ \mbox{}\\
   \displaystyle{ X_2= v \partial_{v} + 2n \partial_{n} + (\tau + 2nv/a)
\partial_{\tau} + (2n-v^2/a) \partial_{\chi} \,, \quad X_3=
   \partial_{v}\,,}\\ \mbox{}\\ \displaystyle{ X_4= n \partial_{\tau} - v
\partial_{\chi} \,, \quad X_5=  (v/a) \partial_{\tau} + \ln n
\partial_{\chi} \,, \quad X_6=  \partial_{\tau}\,, \quad X_7=
\partial_{\chi}. }
\end{array}
$$
The restriction of this group on the solution of
the BVP with the above-mentioned boundary conditions leads to the
three-parameter RG
$$ R_1= X_{1}, \quad R_2= X_{2}, \quad R_3= a X_{3}+X_{4}\,.  $$
As in the previous case, the exact solution of the BVP $\tau=nw$,
$\chi= n-aw^2/2$ appears as an intersection of all invariant manifolds
that correspond to these RG-operators.
\par The characteristic feature of the
described approach is the formulation of boundary data in the form of a
differential constraint and the subsequent search of the group admitted by
this constraint and basic equations. It is evident that there exists an
infinite number of other differential constraints that adequately describe
the same boundary data and the use of which leads to different RG algebras.
As an example, we can point to differential constraints that arise from the
zero equality of appropriate coordinates of the infinite L-B algebra.
\par
The example of RG-symmetry constructing on the basis of
L-B symmetry reveals the practical importance of the latter
and, on the other hand, demonstrates point symmetries that are not
admitted by the original equation. The procedure of construction of
RG-symmetries with the help of a differential constraint was also carried
out in~\cite{KKP_PF13_95} for the linear parabolic equation.

\section{RG as a subgroup of an approximate symmetry group}
An attractive method of RG constructing is that based on approximate
symmetries~\cite{BGI}.  This method can be applied to systems
described in terms of models based on DEs with small
parameters.  These small parameters allows us to consider a simple subsystem
of the original DEs that usually admits an extended symmetry group
inherited by the original DEs.  Restricting this
approximate group on the solution of the BVP  yields the desired
RG-symmetries. The merits of the described method is illustrated below
for the BVP (\ref{eq2.1}) with a small nonlinearity parameter $a \ll 1$.
In terms of the variable $w=v/a$ the following basic system of linear
DEs is obtained instead of (\ref{eq2.2}):
\be \tau_w - (n/\varphi(n))\chi_n = 0\,,
\quad \chi_w + a \tau_n = 0 \,.
\label{eq7.1} \ee
\par
For $a=0$, it admits an infinite L-B symmetry group
\be
X=f\partial_{\tau} + g\partial_{\chi}\,,
\label{eq7.2} \ee
characterized by an arbitrary dependence of the zero-order
coordinates $f=f^0$ and $g=g^0$ upon $n,\, \tau,\, \chi$
and the derivatives ${\tilde{\tau}}_s$, $\chi_s$ of an arbitrary order
\be
\begin{array}{l}
\displaystyle{
f^0= F^0 + \int dw \left\{ (n/\varphi) Y g^0 \right\}, \quad
g^0= G^0 }\,.
\end{array}
\label{eq7.3} \ee
Here and below in (\ref{eq7.5})
$$
\begin{array}{c}
\displaystyle{
Y= \partial_n + \sum\limits_{s=0}^{\infty}\left (
\tau_{s+1} \partial_{\tau_s} + \chi_{s+1}\partial_{\chi_s} \right) \,, } \\
\mbox{} \\
\displaystyle{
\tau_s = \frac{ \partial^s \tau}{ \partial n^s }\,, \quad
\chi_s = \frac{ \partial^s \chi }{\partial n^s }\,, \quad
\tilde{\tau}_s = \tau_s - w
\sum_{p=0}^{s} {s \choose p}
\frac{\partial^p ( n / \varphi)}{\partial n^p } \chi_{s-p+1}
\, ,}
\end{array}
$$
$F^i(n,\chi_s,\tilde{\tau}_s)$ and $G^i(n,\chi_s,\tilde{\tau}_s)$
are arbitrary functions of their arguments, and expressions in curly brackets
before integrating over $w$ should be given in terms of
$\tilde{\tau}_s,\, \chi_s, \, n,\, w$.
\par
For  $0 < a \ll 1$, this symmetry is inherited as an approximate one  by
equations (\ref{eq7.1}) which thus represent an approximate RG-manifold.
For example, for $\varphi (n) =1$ the following result is obtained:
\be
\begin{array}{c}
\displaystyle{
f^i= F^i + \int dw \left\{ Z  f^{i-1}
         + \frac{n}{\varphi} Y g^i \right\}, \qquad
g^i= G^i +  \int dw \left\{Z g^{i-1} -  Y f^{i-1} \right\}\,,
}\\
\mbox{} \\
\displaystyle{
Z=\sum\limits_{s=0}^{\infty}  \tau_{s+1} \partial_{\chi_s} \,, \qquad
\tilde{\tau}_s = \tau_s - w ( n{\chi}_{s+1} + s{\chi}_{s}) \,, \quad
i \geq 1 \,. }
\end{array}
\label{eq7.5} \ee
\par
One can see, that the symmetry of equations (\ref{eq7.1})
for $a=0$ is inherited by the symmetry of these equations
for $a\neq 0$ up to an arbitrary order of this parameter.
It should be noticed that both zero-order and higher-order
 approximate symmetries may appear as Lie point
symmetries or L-B symmetries, and this parameter may be involved
in group transformations, as well.
\par
The restriction of the approximate group obtained on a particular solution
of the BVP defines the specific form of the zero-order symmetries.
It means that while constructing RG-symmetries for the BVP
(\ref{eq7.1}) in view of the bounary data from (\ref{eq2.1}),
coordinates $f^0,\, g^0$  and  "integration constants"
$F^i,\, G^i,\, i\geq 1$ are not arbitrary functions,
but should be chosen so that relations $f=0$, $g=0$
satisfy desired boundary conditions $\tau_s =0,\ \chi =H(n)$ at $w=0$.
Provided that the functions $F^i$ and $G^i,\, i\geq 1$ are also equal to zero
in this case, boundary conditions are correlated with the form of
functions $f^0$ and $g^0$. In general, invariance conditions $f=0$ and $g=0$
appear as differential constraints (or algebraic relations)
to be fitted by boundary data.
\par
Of special interest are such zero-order functions
 $f^0$ and $g^0$ for which infinite series (\ref{eq7.5})
are truncated for some finite value of $i=i_{max}$, and we arrive at
finite sums. In this case, instead of an approximate group with respect to
a small parameter $a$ we obtain the exact symmetry group (compare
with~\cite[\S 11]{BGI}). A simple example of this is given by the
RG-operator (\ref{eq4.6}). It is easily checked that in
terms of $n$ and $w$, the combinations of coordinates $f_3+2(f_7 + f_9)$ and
$g_3+2(g_7 + g_9)$ are expressed as binomial in $a$,
i.e.  expressions for $f$ and $g$ are represented as zero-order and
first-order terms $f=f^0+af^1$
and $g=g^0+ag^1$, where $f^0, \ g^0$ and $f^1, \ g^1$
according to (\ref{eq7.3}) and (\ref{eq7.5}) are defined by the
formulas
\be
\begin{array}{c}
\displaystyle{
f^0 = 2n(1-n)\tau_2 - n\tau_1  -2nw(\chi_1 + n \chi_2)\,, }\\
\mbox{} \\
\displaystyle{
\quad g^0 = 2n(1-n)\chi_2 + (2-3n)\chi_1 \,, } \\
\mbox{}  \\
\displaystyle{
f^1 = \frac{1}{2}nw^2\tau_2 \,, \quad
g^1 = 2nw\tau_2 + w \tau_1 +\frac{1}{2}(nw^2 \chi_2+w^2 \chi_1) \,. }
\end{array}
\label{eq7.6} \ee
>From here, in view of (\ref{eq7.5}), it follows that higher-order
corrections vanish, and we obtain an exact second-order L-B
symmetry of DEs (\ref{eq7.1}) at $\varphi = 1$ for arbitrary $a\neq 0$; this
symmetry gives rise to the exact solution~\cite{KPS_DE_93,K_Pisa-95}
satisfying the boundary condition $N=\cosh^{-2}(x)$ defined by the zero
order term $g^0$.
\par
The arbitrariness in functions $f^0,\, g^0$ enables us to construct
RG-symmetries for any boundary conditions.  As an illustration, we present
RG-symmetries for the BVP  with
\be H(n)=(\ln (1/n))^{1/2} \, ,
\label{eq7.7} \ee
describing space evolution (self-focusing) of the gaussian beam with the
originally plane phase front at $\tau=0$.  To satisfy the initial
distribution (\ref{eq7.7}), one can choose the following functions
$f^0=1+ 2n\chi \chi_1$ and $g^0=0$.
For this value of $f^0$ the inherited point group of the BVP is constructed
with the help of formulas (\ref{eq7.5}) and is given by the
operator
\be R= -2\chi \partial_w + 2a\tau \partial_n +\left( 1+
\frac{a\tau^2}{n} \right) \partial_{\tau}  \,.
\label{eq7.9} \ee
\par
The invariance condition for the solution of the BVP with respect to RG with
this operator is presented in the form of two partial DEs
$$ \chi \chi_w - a\tau \chi_n =0 \,, \quad
     2\chi \tau_w - 2a\tau \tau_n + 1+ (a\tau^2/n) = 0 \,,
$$
the solution of which yields the desired approximate analytical solution
of the problem
\be x^2=(ant^2-\ln n)\left[ 1-2 Q(\sqrt{ant^2})\right]^2\,, \quad
v=-2\frac{x}{t}\frac{Q(\sqrt{ant^2})}{\left[ 1-2 Q(\sqrt{ant^2}) \right]} \,.
\label{eq7.11} \ee
Here the function $Q(z)$ is expressed as follows
$$
Q(z)=z e^{-z^2 /2}\int\limits_{0}^{z} dt e^{t^2 /2}\,.
$$
The first-order approximate symmetry obtained can be used to calculate
a higher-order approximation of the RG-operator (\ref{eq7.9}) and, thus,
to improve the analytical solution (\ref{eq7.11}). One can also obtain
new-type RG-operators just by substituting the approximate solution
(\ref{eq7.11}) into formulas (\ref{eq3.17}).

\section{Conclusion}
\par
This paper presents a new approach to constructing RG-symmetries
based on the mathematical apparatus of classical and modern group
analysis. It differs from the traditionally used methods of
constructing RGs in theoretical physics and is formulated
as a sequence of the following steps:
\par
I) {\it constructing the RG-manifold}, that takes into account both
basic equations and the corresponding boundary conditions;
\par
II) {\it calculating the symmetry group}, admitted by RG-manifold;
\par
III) {\it restricting the group} obtained {\it on the solution of the BVP};
\par
IV) utilizing of RG-operators to {\it find analytical expressions for
solutions}.
\par
As it was shown there exists a set of different algorithms for
finding RG-symmetries. The choice of a particular one for a given
physical problem depends on a mathematical model used for the problem
description. \par
       It should be noted, that different methods of constructing
RG-symmet\-ri\-es described above do not exhaust the suggested approach
(see, e.g., \cite{K_TMP_97,KS_PD41_97}).
Procedure of constructing RG-sym\-met\-ri\-es may combine different
algorithms; for example, of interest is a simultaneous use of the
method based on approximate symmetries and the invariant embedding
method, and so on.

Our approach reveals a close relation of {\it functional
self-similarity} property (i.e., "classical" RG--symmetry as an exact
property of a solution) to an invariance condition of a BVP solution
with respect to RG-operator. Mathematically, the latter is formulated
as the vanishing condition for the coordinate of a canonical RG-operator on
a solution of BVP. \par
       One can readily see that RG-operators may appear in the form,
that is different from QFT case\cite{Book}, e.g., operators
of Lie-B\"acklund RG-symmetries.  However, in some cases "our"
RG-operators can look like that ones in QFT renormalization group.
For example, linear combination of operators $\alpha_1 X_1$ and
$\alpha_3 (X_3 + X_4)$ for the
BVP (\ref{eq5.1}) with $\alpha_1=1$, $\alpha_3=-a$ gives
\be
R=\partial_t  - ax^2 \partial_x, \label{eq8.1}
\ee
which is formally equivalent (with appropriate change of variables
$t=\ln x$, $x=g$ and $\beta (g) = a g^2$) to the differential
operator for one-coupling massless QFT model in one-loop approximation.
\par
 Up to now this approach is feasible for systems that can be described
by DEs and is based on the formalism of modern group analysis.
\par
   It seems also possible to extend our approach on physical systems
that are not described just by differential equations. A chance of
such extension is based on recent advances in group analysis of systems
of integro-differential equations~\cite{Mel,KKP_DE_93} that allow
transformations of both dynamical variables and functionals of a
solution to be formulated~\cite{KKP_JNMP_96}. More intriguing is the
issue of a possibility of constructing a regular approach for more
complicated systems, in particular to that ones having an infinite
number of degrees of freedom. The formers can be represented in a
compact form by functional integrals (or {\it path integrals}).
\par
This work was supported by Russian Foundation for Fundamental Research
(project No 96-01-00195 and partially projects Nos 96-01-01297 and
 96-15-96030).


\begin{references}
\bibitem{ftn1}
The generalization to the case of several parameters is
straightforward -- see, e.g.,\cite{BSh_NC_56,Sh_IJMP_88}.
\bibitem{ftn2}
    In general this means that $S$ transforms according to
    some representation of a group.
\bibitem{ftn4}
{A disadvantage of this is that some examples have a methodological, rather
than a physical significance.}
\end{references}
\end{document}